\documentclass[
superscriptaddress,
 amsmath,
 amssymb,
 aps,
 prl,
 reprint,
]{revtex4-1}

\usepackage{graphicx}
\usepackage{dcolumn}
\usepackage{bm}
\usepackage{color}

\newcommand{\Fig}[1]{Fig.~\ref{#1}}
 
\newcommand{\Eq}[1]{Eq.~(\ref{#1})} 
\newcommand{\Eqs}[2]{Eqs.~(\ref{#1})~and~(\ref{#2})} 
 
\newcommand{\um}{\mu\textrm{m}}
\newcommand{\minus}[1]{\textrm{-}{#1}}

\begin{document}

\title{Aloof electron probing of in-plane SPV charge distributions on GaAs surfaces}

\author{Zilin~Chen}
\affiliation{Department of Physics and Astronomy, Northwestern University, Evanston, Illinois 60208, USA}
\affiliation{Department of Physics and Astronomy, University of Nebraska-Lincoln, Lincoln, Nebraska 68588, USA}

\author{Wayne~Cheng-Wei~Huang}
\email{Email: waynehuang@mx.nthu.edu.tw} 
\affiliation{Department of Physics, National Tsing Hua University, Hsinchu 30013, Taiwan (R.O.C.)}
\affiliation{Department of Physics and Astronomy, University of Nebraska-Lincoln, Lincoln, Nebraska 68588, USA}

\author{Herman~Batelaan}
\email{Email: hbatelaan2@unl.edu} 
\affiliation{Department of Physics and Astronomy, University of Nebraska-Lincoln, Lincoln, Nebraska 68588, USA}

\begin{abstract}
The motion of free electrons moving parallel and above a semiconductor surface can be influenced by shining laser light onto the surface. Here we report strong deflection of aloof electrons by an undoped GaAs surface illuminated with a 633\,nm laser. The deflecting electric field from the surface photovoltaic charges extends 100\,$\um$ into the vacuum. As surface photovoltage (SPV) is sensitive to the electronic states of the GaAs surface, the aloof electron beam serves as a probe for SPV charge dynamics at the mesoscopic length scale. The observed in-plane SPV charge distribution persists beyond 1 second after the laser beam is blocked. Our work suggests the possibility of writing designed 2D charge patterns on semiconductor surfaces with a scanning laser beam, providing unusual flexibility for electron beam manipulation.
\end{abstract}

\maketitle

\section{I. Introduction}

The surface photovoltage (SPV) can facilitate electrostatic near fields above optically illuminated semiconductor surfaces. The near fields are important for SPV spectroscopy \cite{Kronik1999}, for understanding charge separation and recombination processes in solar cells \cite{Chen2024}, for electron beam manipulation \cite{Kampherbeek2000}, and for studying quantum decoherence and dissipation of free electrons \cite{Anglin1997, Caldeira1983}. In this study, we probe the SPV near fields on an undoped single-crystalline GaAs (110) surface through the deflection of aloof electrons. We observe strong deflection when the surface is under superband or subband photoexcitation. As SPV near fields depend on the transport properties and electronic states of the surface, our approach serves as a probe for SPV charge dynamics. 

Recent developments on SPV measurement include SPV microscopy \cite{Chen2024} and scanning ultrafast electron microscopy \cite{Najafi2019}. These methods are direct probes of the 2D photovoltaic charge distribution at the microscopic scale. Our approach complements the existing methods in that we measure the 3D SPV near fields at the mesoscopic length scale. Specifically, the vertical electron deflection due to SPV near fields is proportional to the 1D integral of the SPV charge distribution along the direction of electron beam propagation \cite{Huang2014}. The vertical deflection we report here is up to 200\,$\um$ (7 beam diameters) at 40\,cm after the GaAs surface. It can thus be used with an aperture to switch an electron beam on and off. Also, the SPV near field extends 100\,$\um$ into the vacuum, providing a long working distance. Our method is robust, requires no nanofabrication and works at modest vacuum ($10^{\minus{6}}$\,Torr).

The manipulation of free electrons by laser-illuminated material structures is of general interest and many examples exist for such technology. For example, the electron beam, and many copies thereof, can be steered by laser light in the presence of a surface and used for multibeam electron lithography \cite{Kampherbeek2000}. In addition, control of electron motional states in dielectric laser accelerators \cite{England2014}, control of attosecond electron dynamics near a nanotip \cite{Kruger2011, Nabben2023}, and laser-induced phase modulation of an electron wave \cite{Feist2015} are but a few examples of exquisite motion control through photoinduced near fields. As SPV charge distributions is closely linked to the intensity profile of the laser light, simulation can be used to guide the development of a designed near field structure, providing another useful tool for optical control of free electrons.

A detailed understanding of photoinduced near fields and the ensuing change of surface resistivity are also needed for attaining controlled electron-surface decoherence \cite{Sonnentag2007, Kerker2020, Beierle2018, Chen2020}. This approach was first proposed by Zurek \cite{Anglin1997}, intended for testing Caldeira and Leggett’s quantum dissipation theory \cite{Caldeira1983}. In our experiment, we found that the electron diffraction pattern can be strongly distorted by the gradient force of the SPV near field, but the beam coherence is not affected by the light-modulated surface resistivity.

\begin{figure*}[t]
   \includegraphics[width = \linewidth]{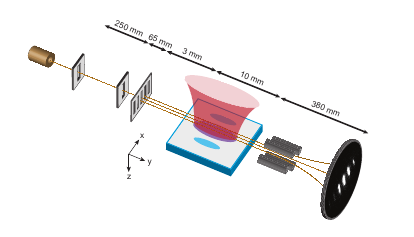} 
  \caption{Schematic of the electron deflection experiment. A thermionic electron gun (top left) emits electrons at an energy of 1.67\,keV. Two collimation slits separated by 250\,mm limit the transverse momentum spread of the beam and deliver an electron beam with a transverse spatial coherence of $\sim 500$\,nm to a nanofabricated grating. The grating has a periodicity of 100\,nm and transmits 50$\%$ of the electron beam. The diffracted electron beam passes parallel to and over a 10\,mm long undoped GaAs surface. The surface is illuminated with an elliptically shaped laser beam. The laser beam widths parallel and transverse to the direction of electron beam propagation are 5\,mm and 100\,$\um$, respectively. The laser wavelength is 633\,nm for superband excitation or 1064\,nm for subband excitation. The nominal laser power is 1\,mW. The laser produces a SPV charge distribution (purple and blue shades) whose electrostatic near field deflects the free electrons at an electron-surface distance up to 100\,$\um$. The deflected electron beam is magnified with a quadrupole lens and recorded with an imaging detector (bottom right).}
\label{fig:setup}
\end{figure*}

\begin{figure}[t]
   \includegraphics[width = \linewidth]{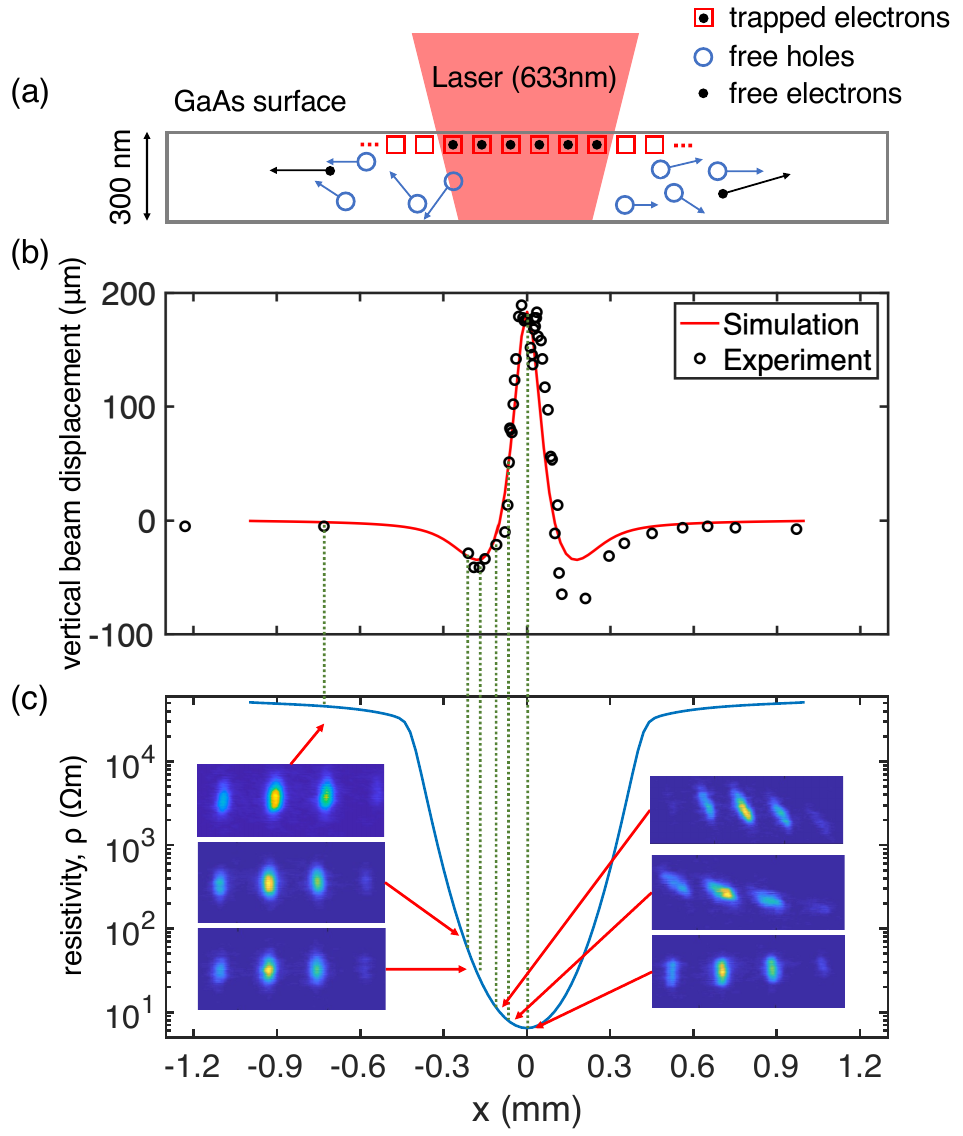} 
  \caption{Superband photoexcitation on the GaAs surface. (a) A schematic of superband excitation. (b) The vertical displacement of a diffracted electron beam is measured (black circles) as the laser beam is moved along the x-axis. Here, the electron-surface distance is 13\,$\um$. In the central region, the deflection is away from the surface due to the repulsive force from the trapped electrons. The superband model together with an electron trajectory simulation (red solid line) gives good agreement with the experiment. (c) The superband model predicts a spatially modulated resistivity (blue solid line) with a four-orders-of-magnitude variation across the laser-illuminated area. The electron diffraction patterns exhibit varying degrees of rotation due to the local field gradient, but the contrast remains unchanged.}
\label{fig:mechanism1}
\end{figure}

 \begin{figure}[t]
   \includegraphics[width = \linewidth]{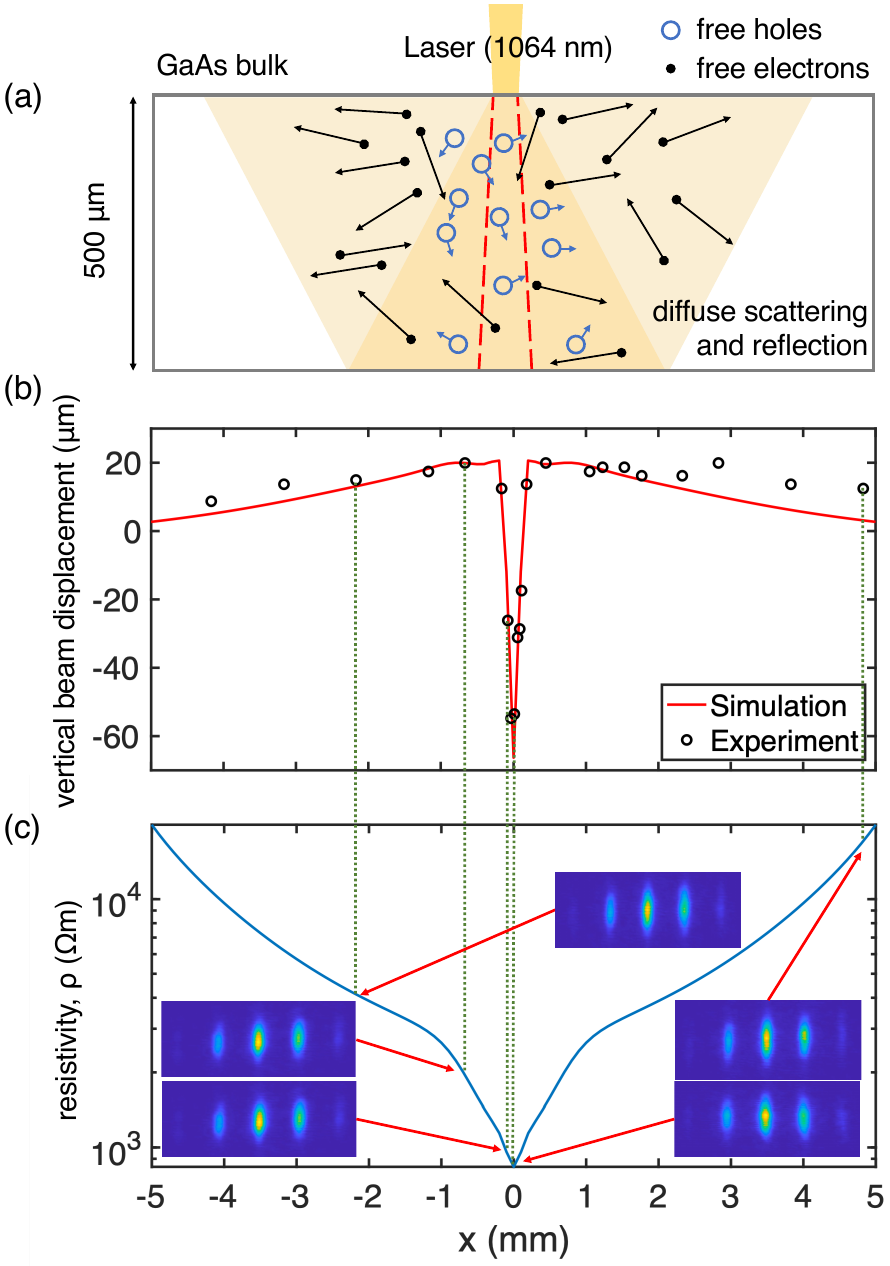} 
   \caption{Subband photoexcitation in the GaAs bulk. (a) A schematic of subband excitation. The 1064\,nm laser with a long penetration depth create free electrons and holes in the bulk. Diffuse scattering and reflection from the back side (yellow shades) extend the carrier generation area to millimeters. (b) The vertical displacement of a diffracted electron beam is measured (black circles) as the laser beam is moved along the x-axis. Comparison between the experiment and the subband model (red solid line) shows good agreement. In the central region, the deflection is towards the surface due to the attractive force from the free holes. Note that free electrons diffuse throughout the full length of the GaAs sample (10\,mm). (c) The laser-modulated surface resistivity (blue solid line) is given by the subband model. The electron diffraction patterns only exhibit mild distortion because the near-field gradient is weak.}
   \label{fig:mechanism2}
\end{figure}

\section{II. Results and discussions}

\subsection{A. Superband Photoexcitation}

In our experiment (\Fig{fig:setup}), in-plane SPV charge distributions are probed with a diffracted electron beam through vertical beam displacement. We use rate equations to model the photovoltaic carrier dynamics \cite{Kronik1999, Liu1993} and an electron trajectory simulation \cite{Chen2020} to compute the near-field interaction that leads to the beam displacement.

As the undoped GaAs has a bandgap of 1.42\,eV (873\,nm), illuminating the surface with a 633\,nm continuous-wave He-Ne laser creates free electrons and holes through superband excitation. The generated electrons and holes are close to the surface because the laser penetration depth in this case is only 300\,nm \cite{Sturge1963}. This results in most free electrons being trapped in the surface states, while the free holes diffuse to the surrounding areas (Fig.~\ref{fig:mechanism1}(a)). The direction of electron beam displacement (Fig.~\ref{fig:mechanism1}(b)) indicates that trapped electrons locate at the center of the charge distribution, while the side wings are populated with free holes. In the equilibrium state, with the laser continuously on, the internal electric field within the charge distribution becomes strong enough to keep the holes from diffusing further away. The magnitude of electron deflection saturates at low laser power (1\,mW) and is independent of laser polarization. 

We use rate equations \cite{Kronik1999, Liu1993, Hook1991} to obtain the equilibrium density distributions for free electron $n(x)$, free hole $p(x)$, and trapped electron $n_{t}(x)$,

\begin{equation}
\frac{dn(x)}{dt}=\frac{1}{e} \frac{dJ_{n}(x)}{dx}+G_{n}(x)-R_{n}(x),
\label{eqn:conduction}
\end{equation}

\begin{equation}
\frac{dp(x)}{dt}=-\frac{1}{e}\frac{dJ_{p}(x)}{dx}+G_{p}(x)-R_{p}(x),
\label{eqn:valence}
\end{equation}

\begin{equation}
\frac{dn_{t}(x)}{dt}=G_{t}(x)-R_{t}(x),
\label{eqn:trap}
\end{equation}
where $G_{i}(x)$ and $R_{i}(x)$ denote the generation and recombination rates, respectively, and $e$ is the electron charge unit. The charge current density $J_{i}(x)$ depends on the balance between diffusion and the drift caused by the internal electric field $E(x)$,

\begin{equation}
J_{n}(x)=e\mu_{n} n(x)E(x)+eD_{n}\frac{dn(x)}{dx},
\end{equation}
\begin{equation}
J_{p}(x)=e\mu_{p} p(x)E(x)-eD_{p}\frac{dp(x)}{dx},
\end{equation}
where $\mu_{n}=8000$\,$\mathrm{cm^2/Vs}$ and $\mu_{p}=300$\,$\mathrm{cm^{2}/Vs}$ are the mobilities of free electrons and holes in undoped GaAs. The corresponding diffusion coefficients are $D_{n}=200$\,$\mathrm{cm^{2}/s}$ and $D_{p}=10$\,$\mathrm{cm^{2}/s}$. The internal electric field $E(x)$ is determiend by the total charge density $\rho(x) = e\left( p(x)-n(x)-n_{t}(x) \right)$. The thermal recombination rate $R_{n}(x)=r_{n\rightarrow p}^{th}+r_{n\rightarrow t}^{th}$ in \Eq{eqn:conduction} is the sum of the thermal transition from the conduction band to the valence band $r_{n\rightarrow p}^{th}=c_{n\rightarrow p}^{th}n(x)p(x)$ and the thermal capture rate from the conduction band to the surface trapping states $r_{n\rightarrow t}^{th}=c_{n\rightarrow t}^{th}n(x)\left(N_{t}-n_{t}(x)\right)$, where $N_{t}$ is the trapping state density \cite{Kronik1999}. On the other hand, the generation rate $G_{n}(x)=g_{p\rightarrow n}^{th}+g_{p\rightarrow n}^{opt}+g_{t\rightarrow n}^{th}+g_{t\rightarrow n}^{opt}$ has four contributions. The $g_{p\rightarrow n}^{th}$ and $g_{p\rightarrow n}^{opt}$ terms correspond to thermal and optical excitation from the valence band to the conduction band. Similarly, the $g_{t\rightarrow n}^{th}$ and $g_{t\rightarrow n}^{opt}$ terms characterize thermal and optical excitation rates from the trapping states to the conduction band. These four terms are given by

\begin{subequations}
\begin{eqnarray}
&g_{p\rightarrow n}^{th}=c_{n\rightarrow p}^{th}n_{0}p_{0},\\
&g_{p\rightarrow n}^{opt}=F(1-R)\eta e^{-x^{2}/\omega_x^{2}} ,\\
&g_{t\rightarrow n}^{th}=\sigma_{t\rightarrow n}^{th}n_{t}(x),\\
&g_{t\rightarrow n}^{opt}=F\sigma_{t\rightarrow n}^{opt}n_{t}(x),
\end{eqnarray}
\end{subequations}
where $n_{0}$ and $p_{0}$ are the intrinsic carrier densities ($\sim$ $10^{6}\,\mathrm{cm^{\minus{3}}}$) in the absence of laser illumination, $F$ is the laser photon flux ($\sim 10^{18}\,\mathrm{cm^{\minus{2}}s^{\minus{1}}}$), $R=0.3$ is the GaAs reflectivity at 633\,nm \cite{Philipp1963}, $\eta=0.7$ is the quantum efficiency \cite{Milanova1999}, $\omega_{x}=100$\,$\um$ is the laser beam width transverse to the direction of electron beam propagation, $\sigma_{t\rightarrow n}^{th}$ and $\sigma_{t\rightarrow n}^{opt}$ are thermal and optical excitation rates from the trapping states to the conduction band. In \Eqs{eqn:valence}{eqn:trap}, the corresponding generation and recombination rates are

\begin{subequations}
\begin{eqnarray}
&G_{p}(x)=g_{p\rightarrow n}^{th}+g_{p\rightarrow n}^{opt},\\
&R_{p}(x)=r_{n\rightarrow p}^{th},\\
&G_{t}(x)=r_{n\rightarrow t}^{th},\\
&R_{t}(x)=g_{t\rightarrow n}^{th}+g_{t\rightarrow n}^{opt}.
\end{eqnarray}
\end{subequations}

We obtain the total charge density $\rho(x)$ by solving the coupled rate equations Eqs.~(\ref{eqn:conduction})-(\ref{eqn:trap}) with parameters given above. To compare the SPV charge distribution $\rho(x)$ with the electron deflection data, we construct an electron trajectory simulation \cite{Chen2020} using an electrostatic field derived from the charge distribution. In Fig.~\ref{fig:mechanism1}(b), we find a good agreement between our model and the deflection data.

The superband model also predicts slow relaxation of the SPV charge distribution after the laser beam is blocked. While the SPV charge distribution is usually established in microseconds or faster \cite{Huang2014, Najafi2019}, its relaxation can take seconds due to surface trapping states \cite{Liu1993, Lagowski1992}. After blocking the laser with a mechanical chopper, we observe that the electron deflection persists for 1\,second. This opens an interesting prospect that scanning a laser beam at a rate above 10\,Hz can result in a ``programmable" 2D surface charge pattern. The resulting SPV near field and its gradient may be exploited for building deformable electron-optical elements such as electrostatic lenses or deflectors. 

\subsection{B. Subband photoexcitation}

In order to verify the relation between slow relaxation and surface trapping states, we illuminate the GaAs surface with a 1064\,nm continuous-wave laser. Even though the 1064\,nm photon energy is less than the GaAs bandgap, free electrons and holes can still be generated via EL2 defect-assisted subband photoexcitation \cite{Martin1981, Germanova1987}. As the laser penetration depth at 1064\,nm exceeds the thickness of our GaAs sample (500\,$\um$), the majority of free electrons and holes are generated in the bulk, and surface trapping states can be ignored. This allows us to measure the SPV relaxation time in the absence of surface trapping states. The long penetration depth of 1064\,nm laser is confirmed by observing a red diffuse spot larger than the incident laser beam waist at the back side of the GaAs sample. A simple sketch in Fig.~\ref{fig:mechanism2}(a) illustrates the main idea. The waist of the unobstructed laser beam is symbolically indicated with red dashed lines, while the diffused beam and its reflection are represented by yellow shades.

The deflection data reveals very long side wings of negative charges and a relatively narrow center peak of positive charges (Fig.~\ref{fig:mechanism2}(b)). The non-zero integral of the deflection curve \cite{Huang2014} implies a negative net charge at the surface. Given that charge carriers are generated throughout the bulk and a non-zero-sum charge distribution resides at the surface, the dimension perpendicular to the surface needs to be included in the subband model. We extend the photovoltaic model to include both surface and bulk charge carriers. The free electron densities $n_{s,b}(x)$ associated with the surface and bulk bands are directly coupled through vertical diffusion. The free hole densities $p_{s,b}(x)$ of the surface and bulk bands are also coupled in the same way. Meanwhile, the internal electric field $E(x)$ responsible for the horizontal drift of free charge carriers in each band is determined by the total charge carrier density $p_{s}(x)+p_{b}(x)-n_{s}(x)-n_{b}(x)$. Since surface trapping states are ignored in the subband model, here the generation and recombination rates do not depend on $r_{n\rightarrow t}^{th}$, $g_{t\rightarrow n}^{th}$, and $g_{t\rightarrow n}^{opt}$. Both primary and diffuse laser beams are used in the simulaion for photoexcitation of the surface and bulk bands. A comparison between our model and the electron deflection data shows good agreement (Fig.~\ref{fig:mechanism2}(b)). The broad side wings across the entire sample is a result of higher mobility and stronger diffusion rates of electrons compared to holes. As we modulate the laser intensity with a mechanical chopper, we determine the upper bound of the SPV relaxation time to be 0.6\,ms.

\section{III. Conclusion}

In summary, we perform an electron deflection experiment with an undoped GaAs (110) surface. The surface is optically illuminated by low-power lasers for either superband or subband photoexcitation. We find good quantitative agreement between our experimental results and existing photovoltaic models. For 633\,nm superband excitation, we observe a narrow central region populated with trapped electrons and equally narrow side wings of free holes. The electrostatic near field of SPV charges causes vertical deflection of a diffracted electron beam at an electron-surface distance up to 100\,$\um$. The presence of surface trapping states significantly impedes the electron-hole recombination at the surface, making the lifetime of the in-plane SPV charge distribution exceed over 1\,second. For 1064\,nm subband excitation, where no surface trapping states are involved, the SPV relaxation time is measured to be shorter than 0.6\,ms. It is perhaps interesting to contemplate scanning a laser beam so that a designed 2D charge pattern could be ``written" on the undoped GaAs surface via superband excitation. The gradient of the SPV near field can act as an electrostatic lens for the electron beam passing over the surface. Such ``programmable" electron-optical elements may add an interesting approach to optical control of free electrons.

\section{Acknowledgments}
We thank Prof. H. Ruda for advice on the effect of EL2 states. The funding of this work comes from NSF PHY-2207697. W.C.H. gratefully acknowledges funding support from NSTC 112-2811-M-001-007 and 113-2811-M-007-008. This work was completed utilizing the Holland Computing Center of the University of Nebraska, which receives support from the UNL Office of Research and Economic Development, and the Nebraska Research Initiative. Manufacturing and characterization analysis were performed at the NanoEngineering Research Core Facility (NERCF), which is partially funded by the Nebraska Research Initiative. The research was performed in part in the Nebraska Nanoscale Facility: National Nanotechnology Coordinated Infrastructure and the Nebraska Center for Materials and Nanoscience (and/or NERCF), which are supported by the National Science Foundation under Award ECCS: 2025298, and the Nebraska Research Initiative.

\end{document}